\documentclass[structabstract]{aa}  
\usepackage{graphicx}
\def\arcsec{$^{\prime\prime}$}
\def\arcmin{$^{\prime}$}

\begin{document}

   \title{Detection of diffuse radio emission in the galaxy clusters A800, A910, A1550, and CL 1446+26}
  
    \subtitle{}

   \author{F. Govoni \inst{1},
           C. Ferrari \inst{2},
           L. Feretti \inst{3},
           V. Vacca  \inst{1,4},
           M. Murgia \inst{1},
           G. Giovannini \inst{3,5},
           R. Perley \inst{6}, \and 
           C. Benoist \inst{2} 
          }

  \institute{INAF - Osservatorio Astronomico di Cagliari,
             Strada 54, Loc. Poggio dei Pini, 09012 Capoterra (Ca), Italy
         \and 
             Laboratoire Lagrange, UMR7293, Universit\'e de 
             Nice Sophia-Antipolis, CNRS, Observatoire de la C\^ote d'Azur, 06300 Nice, France
         \and             
             INAF - Istituto di Radioastronomia, Via P.Gobetti 101, 
             40129 Bologna, Italy 
         \and
             Dipartimento di Fisica, Universit\`a degli Studi di Cagliari, Cittadella Universitaria, I-09042 Monserrato (CA), Italy
         \and 
             Dipartimento di Astronomia, Universit\`a degli Studi di Bologna, 
             Via Ranzani 1, 40127 Bologna, Italy
         \and       
              National Radio Astronomy Observatory, Socorro, NM 87801, USA 
            }

   \date{Received September 15, 1996; accepted March 16, 1997}

  \abstract
  {Radio halos are elusive sources located at the center of
   merging galaxy clusters.
   To date, only $\sim$40 radio halos are known, thus the discovery of
   new halos provide important insights on this class of sources.}
  {To improve the statistics of radio halos, we investigated 
   the radio continuum emission in a sample of galaxy clusters.}
  {We analyzed archival Very Large Array observations at 1.4 GHz, with 
   a resolution of $\simeq$ 1\arcmin. 
   These observations complemented by X-ray, optical, and higher 
   resolution radio data allowed to detect a
   new radio halo in the central region of A800 and A1550.
   We discovered a radio relic in the periphery of A910, and finally
   we revealed both a halo and a relic in CL1446+26.}
  {Clusters hosting these new halos show an offset between
   the radio and the X-ray peak. 
   By analyzing this offset statistically 
   we found that radio halos can be quite asymmetric with respect 
   to the X-ray gas distribution, with an average radio$-$X-ray displacement
   of about 180 kpc. When the offsets are normalized by the halo size,
   there is a tendency for smaller halos to show larger displacements.
   }
   {}


   \titlerunning{New large-scale diffuse radio emissions in clusters of galaxies}
   \authorrunning{F. Govoni et al.}
   \maketitle
%

\section{Introduction}
Sensitive radio observations have revealed diffuse emission from the
central regions of some merging galaxy clusters 
(e.g. Giovannini et al. 1999, Giovannini \& Feretti 2000, 
Kempner \& Sarazin 2001, Govoni et al. 2001a, Bacchi et al. 2003, 
Venturi et al. 2007, Venturi et al. 2008, van Weeren et al. 2009, 
Rudnick \& Lemmerman 2009, Giovannini et al. 2009, van Weeren et al. 2011).  
These radio sources, extended over volumes
of $\sim$1 Mpc$^3$ and called radio halos, are diffuse,
low-surface-brightness ($\simeq$ 1$\mu$Jy~arcsec$^{-2}$ at 1.4 GHz),
and steep-spectrum\footnote{$S(\nu)\propto \nu^{- \alpha}$, with
$\alpha$=spectral index} ($\alpha>1$) synchrotron sources with
no obvious optical counterparts (see e.g. Ferrari et al. 2008, 
Feretti et al. 2012).

Radio halos can have quite different length scales, but the largest
halos are the most powerful, in such a way that all these sources may
have similar synchrotron emissivity (Murgia et al. 2009).  
Their radio power at 1.4 GHz correlates with the cluster X-ray
luminosity (Feretti 2002), temperature (Liang 1999, Colafrancesco 1999), 
mass (Govoni et al. 2001a), and Sunyaev-Zel'dovich effect (Basu 2012) 
measurements. 

The radio halo morphology is often very similar to the X-ray-emitting
thermal intra-cluster medium (e.g. Govoni et al. 2001b, 
Feretti et al. 2001, Keshet \& Loeb 2010, Brown \& Rudnick 2011). 
Furthermore, radio halos are preferentially found in clusters that 
show evidence of merger activity (e.g. Buote 2001, Schuecker et al. 2001,
Govoni et al. 2004, Cassano et al. 2010, Rossetti et al. 2011), which
suggests a connection between the origin of radio halos 
and gravitational processes of cluster formation, 
although a one-to-one association between 
merging clusters and radio halos is not supported by present observations. 

Detailed images of radio halos can provide information on the
cluster magnetic field since the halo brightness fluctuations and the
polarization level are strictly related to the intra-cluster magnetic
field power spectrum (Tribble 1991, Murgia et al. 2004, Govoni et al. 2006).  
Vacca et al. (2010) presented a study of the magnetic field power
spectrum in the galaxy cluster A665, which contains a Mpc-scale radio
halo. Their modeling suggests that radio
halos can be effectively polarized, but because of their faintness,
detecting this polarized signal is a very hard task with the current
radio interferometers.  Indeed, radio halos are usually found to be
unpolarized.  Polarized emission from radio halos has been observed so
far only in filaments of the clusters A2255 
(Govoni et al. 2005, see also Pizzo et al. 2011) 
and MACS J0717+3745 (Bonafede et al. 2009).

The statistics of radio halos remain poor because of their faintness.
To date, only $\sim$40 radio halos are known (see e.g. the recent
compilation by Feretti et al. 2012 and 
references therein), thus the discovery of new halos
provide important insights on this class of sources.  As part of an
ongoing program aimed to investigate, in complex X-ray cluster
systems, the presence of halo emission from the sub-clusters in the
merger, we recently discovered the case of the double 
merging system, A399
and A401, which both contain a radio halo and can be considered the
only case so far of a double-radio halo system (Murgia et al. 2010).
We also detected an intriguing diffuse radio emission in A781 (Govoni et
al. 2011), an exceptional system characterized by a complex of several
clusters. The archival VLA data set containing the observation of
A781 includes uniform observations for several galaxy clusters.  
Thus, to improve the scanty statistics of radio halos, we analyzed the entire
data set and the results of this investigation, complemented by X-ray,
optical, and higher resolution radio data are presented in
this work.

The intrinsic parameters quoted in this paper are computed for a 
$\Lambda$CDM cosmology with $H_0$ = 71 km s$^{-1}$Mpc$^{-1}$,
$\Omega_m$ = 0.27, and $\Omega_{\Lambda}$ = 0.73.
Values taken from the literature have been scaled to this cosmology.

\section{VLA Observations and data reduction}

We have analyzed archival observations taken with the VLA at 1.4 GHz
in D configuration (VLA Program AM0469).
This data set contains 41
galaxy clusters observed on average for 15 minutes each.
A brief discussion of the quality of these data is given in the Appendix.

The data were calibrated in phase and amplitude.  
Data editing has been made in order to excise RFI.  
Since the observation was done in spectral line mode,
data were calibrated in bandpass and data editing
has been made channel by channel. 
Surface brightness images were
produced following the standard procedures: Fourier-Transform, Clean,
and Restore implemented in the AIPS task IMAGR. 
We used the Multi-scale CLEAN (see e.g. Greisen et al. 2009), 
an extension of the classical Clean algorithm, 
implemented in the task IMAGR.
We averaged the 2 IFs (and the 7
channels) together in the gridding process under IMAGR.
Self-calibration (phase) was performed to increase the dynamic range
and the sensitivity of the radio images. 
The spectral line mode of this data set is not suitable to produce
polarization sensitive images.

We detected evidence of diffuse radio emission in seven clusters
already known to host a radio halo:
A1758, A851, A1995 (Giovannini et al. 2009); 
A1351 (Giacintucci et al. 2009, Giovannini et al. 2009); 
A773, A2254 (Govoni et al. 2001a); 
A2219 (Bacchi et al. 2003, Orr{\'u} et al. 2007).

In addition, we discovered new diffuse sources in
A800, A910, A1550, CL1446+26, and A781. 
We analyzed the radio continuum emission and 
the spectral index properties between 1.4 and 0.3 GHz of A781
in a separate paper (Govoni et al. 2011). In this paper
we present the properties of the remaining four 
clusters (A800, A910, A1550, CL1446+26), 
for which only data at 1.4 GHz are available 
in the archive. 
We note that a peripheral diffuse emission, 
classified as a radio relic, was previously detected in CL1446+26 
by Giovannini et al. (2009). However, on the basis of the data presented here
the classification of the diffuse emission in this cluster
is more complex than previously stated.

The details of the analyzed radio observations 
(pointing, frequency, bandwidth, VLA configuration,
observing time, observing date, and VLA program) are reported in
Table\,\ref{tabl}. 

To separate the diffuse radio emission from discrete sources, we produced
with standard procedures images at higher resolution obtained 
at 1.4 GHz with the VLA in C configuration for the galaxy 
clusters A800, A910, and A1550.  
In the case of CL1446+26 we used the VLA data
in C configuration previously analyzed by Giovannini et al. (2009).

The position and the flux density\footnote{All the flux densities given in the tables and in the text have been 
primary beam corrected.} of the discrete sources 
embedded in the cluster diffuse emission are 
given in Table\,\ref{tab2}.
These values have been subtracted to the flux density 
obtained by integrating the surface brightness 
of the D array data set down to the 3$\sigma$ level to estimate
the residual flux density of the diffuse emission (see Table\,\ref{radio}).
We note, however, the residual flux density associated with
the diffuse cluster emission must be interpreted with caution.
Indeed, a possible variation in the discrete sources flux density, 
the slightly different frequency of the two data-sets, 
and any absolute calibration error between 
them could result in under or over subtraction of the calculated flux density.
In addition, owing to the short exposure time of the archival observations,
some diffuse emission could be missed. 
Therefore, a deep follow-up investigation would be necessary 
to ensure the recovery of the entire radio flux and to unambiguously 
separate the diffuse emission from that 
of the unrelated discrete sources.

To ensure that the large-scale diffuse emission
is not caused by the blending of discrete sources, for each cluster 
we obtained a total intensity image at 1.4 GHz with 
the VLA in D configuration, after subtraction of discrete sources.
We first produced an image of the discrete 
sources by using only the longest baselines of the 
D configuration data set. 
The clean components of the discrete sources were then subtracted 
from the original data set in the uv-plane by using the AIPS task UVSUB. 
At this points, we calculated the flux density of the residual diffuse emission 
by integrating the surface brightness down to the 3$\sigma$ level, 
and we checked
its consistency with the values reported in Table\,\ref{radio}. 
We found that the agreement is generally good (see the notes on individuals clusters).

\begin{table*}
\caption{Details of the VLA observations.}
\label{tab1}
\centering
\begin{tabular}{lcccccclcr}
\hline\hline
Cluster&  R.A.       & Decl.   &Frequency   & Bandwidth & Conf. & Obs. Time & Date  & Program  \\
       &  (J2000)    & (J2000) & (GHz)      & (MHz)     &       & (min.)   &       &          \\
\hline
A800    & 09 28 22.9 & 37 48 09.0 &  1365/1435 &  7$\times$3.125     &  D & 15    & 1995-March-15 & AM0469  \\
        & 09 25 41.9 & 37 56 50.0 &  1452/1502 &  25                 &  C &  7    & 1984-May-04   & AO0048 \\
A910    & 10 02 57.7 & 67 09 23.0 &  1365/1435 &  7$\times$3.125     &  D & 15    & 1995-March-15 & AM0469  \\
        & 10 03 18.8 & 67 11 02.4 &  1465/1385 &  50                 &  C &  3    & 1993-Jun-28   & AL0297 \\
A1550   & 12 29 19.2 & 47 37 58.0 &  1365/1435 &  7$\times$3.125     &  D & 15    & 1995-March-15 & AM0469 \\
        & 12 28 57.7 & 47 37 58.0 &  1365/1435 &  7$\times$3.125     &  C & 75    & 2003-Jan-03   & AM0702 \\
CL1446+26 & 14 49 28.7 & 26 07 54.1 &  1365/1435 &  7$\times$3.125     &  D & 20    & 1995-March-15 & AM0469     \\
\hline
\multicolumn{9}{l}{\scriptsize Col. 1: Cluster name; Col. 2, Col. 3: Observation pointing 
(R.A. J2000, Decl. J2000); Col. 4: Frequency (IF1/IF2); Col. 5: Bandwidth;}\\
\multicolumn{9}{l}{\scriptsize  Col. 6: VLA configuration;
Col. 7: Observing time (minutes); Col. 8: Observing date; Col. 9: VLA program.}\\
\end{tabular}
\label{tabl}
\end{table*}

\section{New diffuse radio emissions}

For each cluster we analyze the results of the VLA observations with a
particular emphasis on the analysis of the radio properties of the
large-scale cluster diffuse emission. 
 
There are only sparse radio, X-ray, and optical information in the
literature for the clusters analyzed in this paper, and nothing is 
reported regarding their dynamical state. 
In order to test the possible connection between the detection of
diffuse radio emission and the dynamical state of the observed
clusters, we compare our results with the gas and galaxy distribution
of each system. 

We investigated the gaseous environment of these clusters
by checking the X-ray emission in the Rosat
All Sky Survey (RASS). The X-ray images in the 0.1$-$2.4\,{\rm keV}
band were background and exposure corrected, and 
smoothed with a $\sigma$=45$''$ Gaussian kernel.

We produced density maps of the bi-dimensional galaxy distribution of each observed cluster 
on the basis of a multi-scale approach, as described in Ferrari et al. (2005).  
In order to avoid as much as possible projection effects, we selected galaxies 
lying on the cluster red-sequence and brighter than $L^* + 3$.  
For most of the systems of our sample we used the SLOAN Digital Sky Survey
(SDSS) multi-band galaxy 
catalogue (Abazajian et al. 2009), applying a color-magnitude selection 
based on the $(r - i)$ vs. $r$ and $(i - z)$ vs. $r$ plots (see, e.g., Goto et al. 2002). 
Only the field around A910 is not covered by SDSS observations, 
therefore we exploited the SuperCOSMOS catalogue (Hambly et al. 2001) and adopted 
the $(B - R)$ vs. $B$ and $(B - R)$ vs. $R$ plots for color selection.

In the following, for each cluster we present the radio emission 
compared with optical and X-ray images.

\begin{table}
\caption{Information on discrete radio sources.}
\begin{center}
\begin{tabular} {lcccr} 
\hline\hline
Cluster & Label   &  R.A.      &  Decl.       &  S$_{\rm 1.4 GHz}$      \\
        &         &  (J2000)   &  (J2000)     &  (mJy)       \\
\hline
A800    &    A    & 09 28 36.1 & 37 47 06     &  3.5$\pm$ 0.3     \\
        &    B    & 09 28 27.5 & 37 45 12     &  2.3$\pm$ 0.2      \\
\hline
A910    &    A    & 10 03 20.8 & 67 09 32     & 25.2$\pm$1.3    \\
        &    B    & 10 03 15.2 & 67 10 20     &  4.9$\pm$0.3   \\
        &    C    & 10 03 14.1 & 67 11 09     &  1.8$\pm$0.2   \\
        &    D    & 10 03 15.3 & 67 12 00     &  0.7$\pm$0.2   \\
        &    E    & 10 03 24.7 & 67 13 22     &  2.8$\pm$0.3    \\
        &    F    & 10 03 32.2 & 67 11 20     &  2.8$\pm$0.3    \\
\hline
A1550   &    A    & 12 29 02.3 & 47 36 56     & 15.8$\pm$0.8    \\
        &    B    & 12 28 58.5 & 47 38 26     &  2.2$\pm$0.2    \\
        &    C    & 12 29 15.7 & 47 37 04     &  0.8$\pm$0.2    \\
\hline
CL1446+26 &    A  & 14 49 29.7 & 26 07 54     & 29.7$\pm$1.5    \\
          &    B  & 14 49 28.2 & 26 08 22     &  2.4$\pm$0.2   \\
          &    C  & 14 49 30.6 & 26 09 11     &  2.2$\pm$0.2   \\
          &    D  & 14 49 33.1 & 26 04 06     & 11.0$\pm$0.5   \\
          &    E  & 14 49 26.7 & 26 05 21     &  3.2$\pm$0.2   \\
          &    F  & 14 49 18.2 & 26 05 30     &  2.5$\pm$0.2   \\
\hline
\multicolumn{5}{l}{\scriptsize Col. 1: Cluster name; Col. 2: Source label;}\\ 
\multicolumn{5}{l}{\scriptsize Col. 3, 4: Source position (R.A. J2000, Decl. J2000);}\\ 
\multicolumn{5}{l}{\scriptsize Col. 5: Source flux density (from the C array data set.)}\\
\end{tabular}
\label{tab2}
\end{center}
\end{table}

\begin{figure*}[t]
\centering
\includegraphics[width=16 cm]{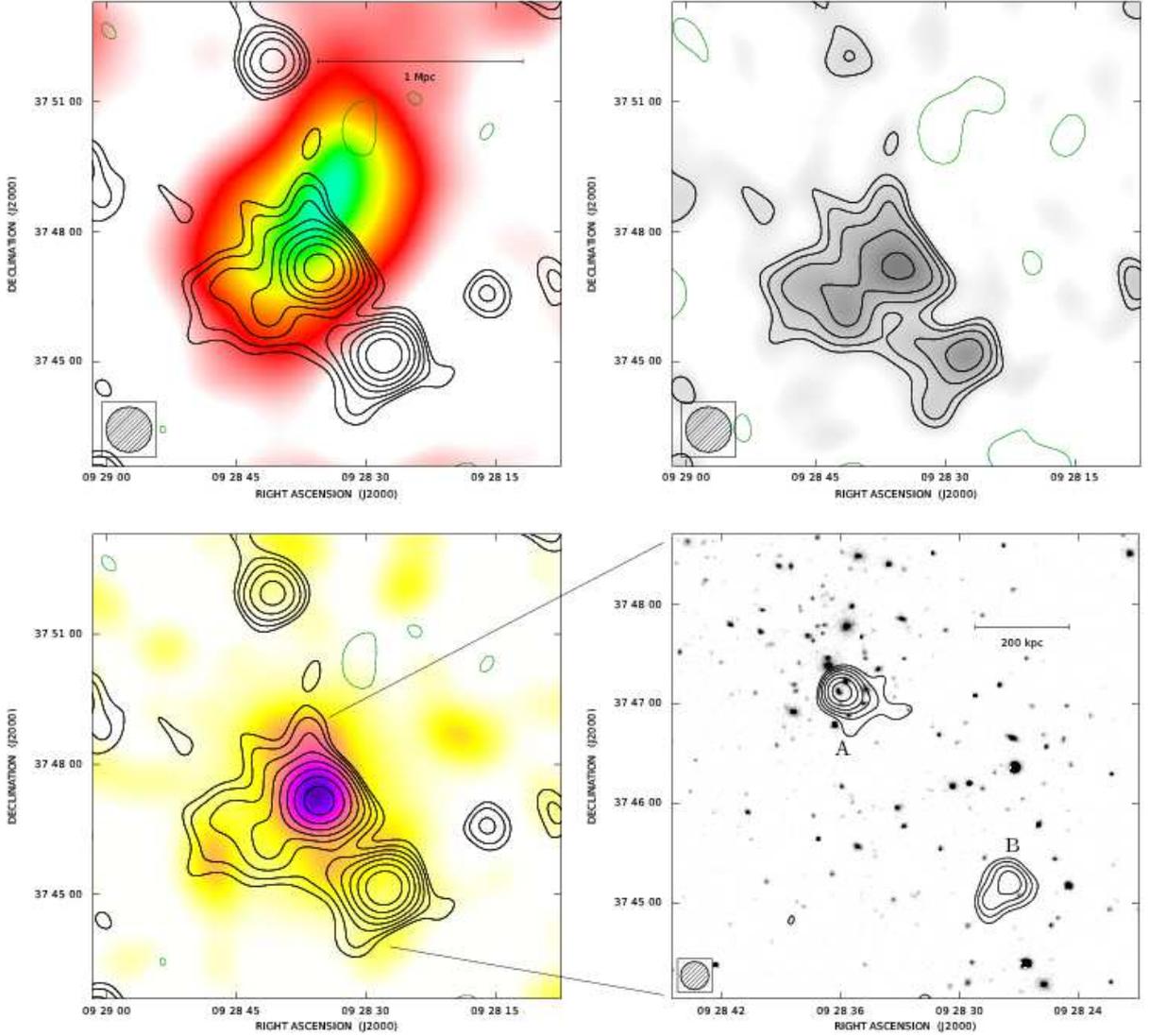}
\caption{{\bf Abell 800.}
  {\it Top, left}: total intensity radio contours of A800 at 1.4
  GHz with the VLA in D configuration. The image has a FWHM of
  63\arcsec$\times$63\arcsec.  The contour
  levels are drawn at $-$0.35 (green), 0.35 mJy/beam, and the rest are
  spaced by a factor of $\sqrt2$.  The sensitivity (1-$\sigma$) is 0.12
  mJy/beam.  Total intensity radio contours are overlaid on the Rosat
  PSPC X-ray image in the 0.1$-$2.4 keV band, taken from the RASS.  
  The X-ray image has been convolved with a Gaussian of
  $\sigma$=45\arcsec.
  {\it Top, right}: total intensity radio contours and grey scale image
  at 1.4 GHz with the VLA in D configuration after subtraction of 
  discrete sources.
  {\it Bottom, left}: total intensity radio contours of
  A800 overlaid on the isodensity map of likely cluster members.
  {\it Bottom, right}: zoom of total intensity radio contours in the 
  center of A800
  at 1.4 GHz with the VLA in C configuration.  The image has a FWHM
  of 17\arcsec$\times$17\arcsec.  The first contour
  level is drawn at 0.3 mJy/beam, and the rest are spaced by a factor
  $\sqrt{2}$.  The sensitivity (1$\sigma$) is 0.1 mJy/beam.  The
  contours of the radio intensity are overlaid on the optical image
  taken from the SDSS (red plate).}
\label{a800}
\end{figure*}

\subsection{Abell 800}

A800 (RXCJ0928.6+3747) is a low X-ray
luminosity cluster at a redshift z=0.2223 which is part of the 
Northern ROSAT All-Sky (NORAS)
Galaxy Cluster Survey by B{\"o}hringer et al. (2000).
Its X-ray luminosity in the 0.1$-$2.4\,{\rm keV}
band is $2.25\times 10^{\rm 44}$\,{\rm erg/sec}.

Low resolution radio contours at 1.4 GHz of A800 are shown in the top-left
and bottom-left panels of Fig.~\ref{a800}. 
This image was obtained with the VLA in D
configuration and has been convolved to a resolution of 
63\arcsec$\times$63\arcsec.
At the top, the radio contours are overlaid on the cluster X-ray
image taken from the RASS. 
At the bottom, the radio contours are overlaid on the isodensity map of
likely cluster members.
The X-ray image shows an elongation in the
northwest-southeast direction. The galaxy distribution has a main clump, 
surrounded by several substructures. 
The diffuse radio emission, that we classify as a radio halo, 
is concentrated within the main galaxy over-densities, 
while a clear shift between the galaxies and gas distribution is present.

To separate the diffuse radio emission from discrete sources we
produced an image at higher resolution. In the bottom-right panel of
Fig.~\ref{a800} we present the radio contours of A800 at 1.4 GHz
taken with the VLA in C configuration. This image has been convolved 
to a resolution of 17\arcsec$\times$17\arcsec.  
The radio contours are overlaid on
the optical image taken from the SDSS.  Two
discrete radio sources (labelled with A and B) are embedded in the
cluster diffuse emission. Source A is located on south of the cluster
X-ray center while it is located in coincidence with the concentration
of likely cluster members. Source B is located on south-west
of the cluster X-ray center at a projected distance of about 2.5$'$ from
source A. 

The total flux density is calculated from the D configuration image 
after a primary beam correction by integrating the total intensity surface
brightness in the region of the diffuse emission down to the 3$\sigma$
level. The resulting total flux density is estimated to be 16.4$\pm$0.8 mJy. 
By subtracting the flux density (see Table\,\ref{tab2}) 
of the embedded discrete sources A and B, a flux density of
10.6$\pm$0.9 mJy appears to be associated with the low brightness
diffuse emission. 
This flux density value corresponds to a radio
power of $P_{\rm 1.4\,GHz} =1.52\times10^{24}$\,{\rm W~Hz}$^{\rm -1}$.

To ensure that the large-scale diffuse emission 
is not caused by the blending of discrete sources, in  
the top-right panel of Fig.~\ref{a800} we present
the total intensity radio contours at 1.4 GHz with the 
VLA in D configuration after subtraction of discrete sources.
In perfect agreement with the value given in Table\,\ref{radio}, 
the flux density of the diffuse emission calculated on the image 
with the point sources subtracted is 10.7$\pm$0.7 mJy.
As measured from the 3$\sigma$ radio
isophote, the overall diffuse emission shows a Largest Linear Size
(LLS) of about 6$'$ ($\simeq 1.28$ Mpc).
However, as pointed out in Murgia et al. (2009), we note that the size
of the diffuse emission calculated from the contour levels should be
considered carefully, since it depends on the sensitivity of the
radio image.

\subsection{Abell 910}

A910 (RXCJ1003.1+6709) is a galaxy cluster at a redshift z=0.2055 which
belongs to the extended sample of the Rosat Brightest Cluster sample 
by Ebeling et al. (2000). Its X-ray luminosity in the 0.1$-$2.4\,{\rm keV}
band is $3.32\times 10^{\rm 44}$\,{\rm erg/sec}. This value is in agreement within the
error with that given by B{\"o}hringer et al. (2000). 

\begin{figure*}[t]
\centering
\includegraphics[width=16 cm]{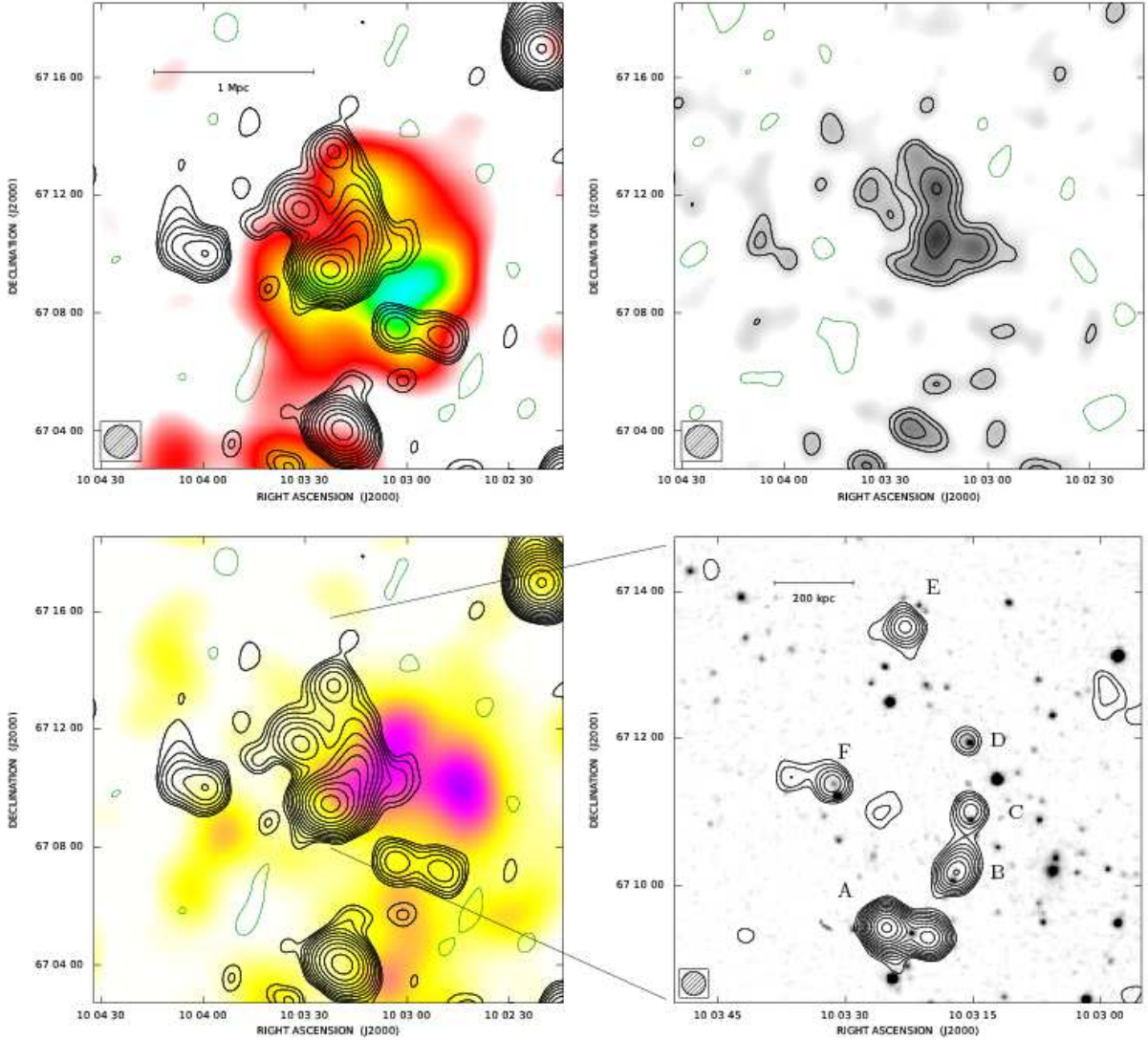}
\caption{{\bf Abell 910.} 
  {\it Top, left}: total intensity radio contours of A910 at 1.4
  GHz with the VLA in D configuration. 
  The image has a FWHM of
  63\arcsec$\times$63\arcsec.  The contour
  levels are drawn at $-$0.45 (green), 0.45 mJy/beam, and the rest are
  spaced by a factor of $\sqrt2$.  The sensitivity (1-$\sigma$) is
  0.15 mJy/beam.  Total intensity radio contours are overlaid on the
  Rosat PSPC X-ray image in the 0.1$-$2.4 keV band, taken from the RASS.  
  The X-ray image has been convolved with a Gaussian
  of $\sigma$=45\arcsec.   
  {\it Top, right}: total intensity radio contours and grey scale image 
  at 1.4 GHz with the VLA in D configuration after subtraction of 
  discrete sources.
  {\it Bottom, left}: total intensity radio contours
  of A910 overlaid on the isodensity map of likely cluster members.
  {\it Bottom, right}: zoom of total intensity radio contours in 
  the center of A910 at 1.4 GHz with the VLA in C configuration.
  The image has a FWHM of  
  18\arcsec$\times$18\arcsec.  The first contour
  level is drawn at 0.3 mJy/beam, and the rest are spaced by a factor
  $\sqrt{2}$.  The sensitivity (1$\sigma$) is 0.1 mJy/beam.  The
  contours of the radio intensity are overlaid on the optical image
  taken from the Digital Sky Survey (red plate).}
\label{a910}
\end{figure*}

The radio contours at 1.4 GHz of A910 are shown in the top-left
and bottom-left panels of Fig.~\ref{a910}.
This image was obtained with the VLA in D
configuration and has been convolved to a resolution of 
63\arcsec$\times$63\arcsec.
At the top, the radio contours are overlaid on the cluster X-ray
image taken from the RASS. 
At the bottom, the radio contours are overlaid on the isodensity map of
likely cluster members.
In A910 at least three X-ray peaks can be identified, possible
indications of a multiple merger. 
The galaxy distribution shows several substructures too.
The diffuse radio emission has been
detected in correspondence of a bridge connecting a bright X-ray
region with a fainter clump. 

In the bottom-right panel of
Fig.~\ref{a910} we present the radio contours of A910 at 1.4 GHz
taken with the VLA in C configuration. This image has been convolved 
to a resolution of 18\arcsec$\times$18\arcsec.  
The radio contours are overlaid on
the optical image taken from the Digital Sky Survey. 
Several discrete radio sources are embedded in the
cluster diffuse emission. 

Note that, in the classical framework of cluster radio source 
classification (e.g. Ferrari et al. 2008, Feretti et al. 2012), 
the diffuse radio source in A910 could be considered as a relic due to its morphology
and the off-set between the radio and both the intra-cluster medium
and the galaxy density peaks.
Indeed, as shown in top-right panel of Fig.~\ref{a800} where the discrete sources 
have been subtracted, the diffuse emission shows an elongated structure
with a LLS of about 5$'$ ($\simeq 1.0$ Mpc) offsetted 
from the main gas density and galaxy density peaks of about 
135$''$ (450 kpc) and about 200$''$ (670 kpc) respectively.
Therefore, given its location and morphology we classify 
this cluster diffuse emission as a relic. 

The total flux density of the relic is calculated
from the D configuration image 
by integrating the total intensity surface
brightness in the region of the diffuse emission down to the 3$\sigma$
level. The resulting total flux density is estimated to 
be 50.3$\pm$2.5 mJy.
By subtracting the flux density of 
the discrete sources A, B, C, D, E, and F (see Table\,\ref{tab2}),
a flux density of 12.1$\pm$2.9 mJy appears to be associated 
with the low brightness relic emission. 
This flux density value corresponds to a radio
power of $P_{\rm 1.4\,GHz} =1.45\times10^{24}$\,{\rm W~Hz}$^{\rm-1}$. 
The flux density calculated on the image with the point sources 
subtracted is 9.0$\pm$0.6 mJy, consistent within the errors with the
value given above.

\subsection{Abell 1550}

\begin{figure*}[t]
\centering
\includegraphics[width=16 cm]{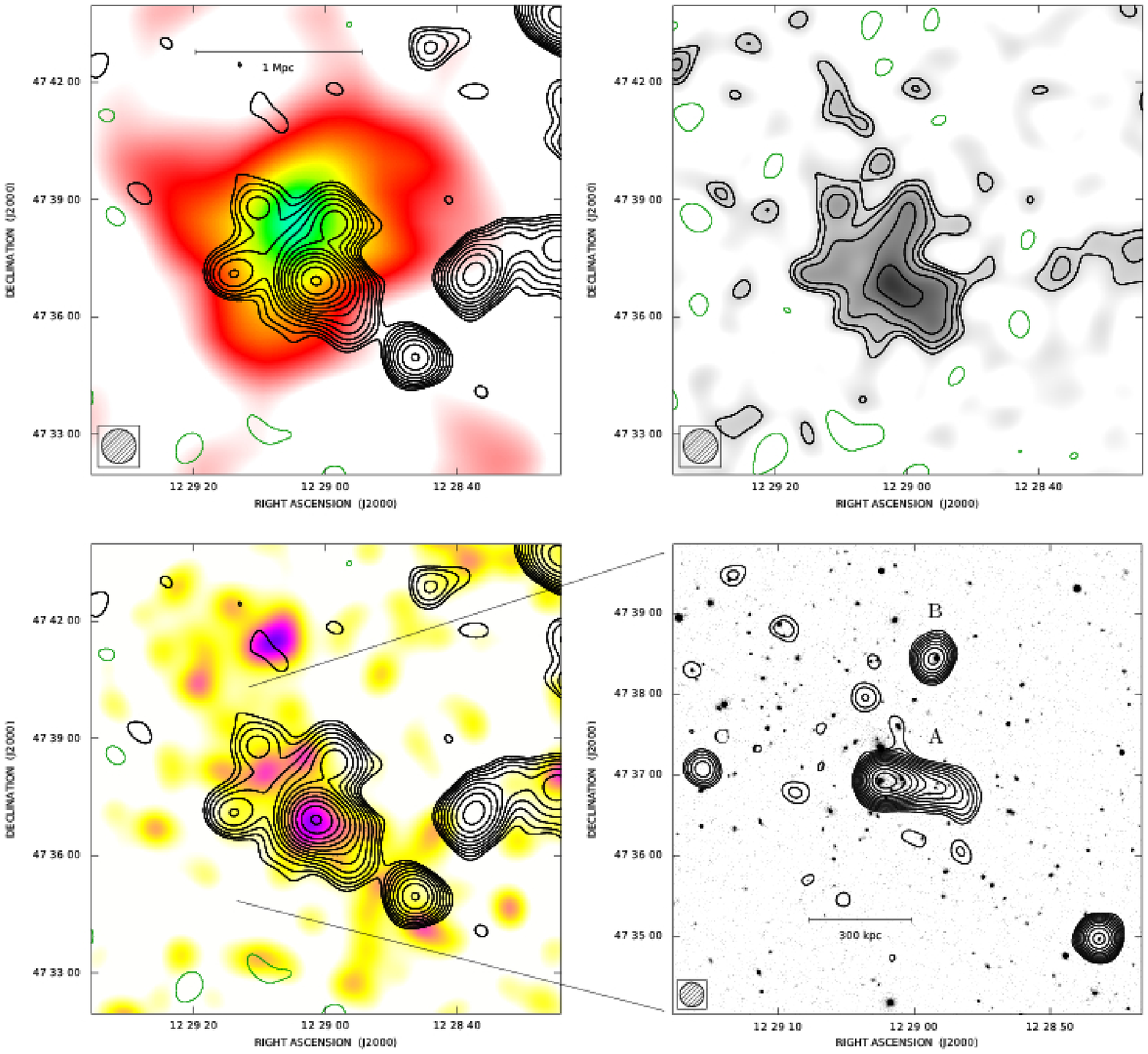}
\caption{{\bf Abell 1550.} 
  {\it Top, left}: total intensity radio contours of A1550 at 1.4
  GHz with the VLA in D configuration. 
  The image has a FWHM of
  53\arcsec$\times$53\arcsec.  The contour
  levels are drawn at $-$0.2, 0.2 mJy/beam, and the rest are spaced by
  a factor of $\sqrt2$.  The sensitivity (1-$\sigma$) is 0.07
  mJy/beam.  Total intensity radio contours are overlaid on the Rosat
  PSPC X-ray image in the 0.1$-$2.4 keV band, taken from the RASS.  
  The X-ray image has been convolved with a Gaussian of
  $\sigma$=45\arcsec.  
  {\it Top, right}: total intensity radio contours and grey scale image 
  at 1.4 GHz with the VLA in D configuration after subtraction of 
  discrete sources.
  {\it Bottom, left}: total intensity radio contours of
  A1550 overlaid on the isodensity map of likely cluster members.
  {\it Bottom, right} : zoom of total intensity radio contours in the center of
  A1550 at 1.4 GHz with the VLA in C configuration. The image has 
  a FWHM of 18\arcsec$\times$18\arcsec.  
  The first contour
  level is drawn at 0.15 mJy/beam, and the rest are spaced by a factor
  $\sqrt{2}$.  The sensitivity (1$\sigma$) is 0.04 mJy/beam.  The
  contours of the radio intensity are overlaid on the optical image
  taken from the SDSS (red plate).}
\label{a1550}
\end{figure*}

A1550 (RXCJ1229.0+4737) is a galaxy cluster at a redshift z=0.2540 with
an X-ray luminosity in the 0.1$-$2.4\,{\rm keV} band 
of $3.51\times 10^{\rm 44}$\,{\rm erg/sec} (B{\"o}hringer
et al. 2000).

Low resolution radio contours at 1.4 GHz of A1550 are shown in the top-left
and bottom-left panels of Fig.~\ref{a1550}.
This image was obtained with the VLA in D
configuration and has been convolved to a resolution of 
53\arcsec$\times$53\arcsec.
At the top, the radio contours are overlaid on the cluster X-ray
image taken from the RASS.
At the bottom, the radio contours are overlaid on the isodensity map of
likely cluster members.  The external X-ray emission is elongated in
the east-west direction, while the inner region is extended along
north-south.  The optical galaxy distribution is bimodal but with an
elongation that however does not follow the cluster X-ray emission.  
Surrounding the X-ray and the optical center we detected a diffuse 
low-surface brightness radio emission, which we classified as a radio halo. 
As in the case of A800, the diffuse radio emission seems to be more spatially 
correlated to the galaxy than to the gas over-densities.

To separate the diffuse radio emission from discrete sources we
produced an image at higher resolution.
In the bottom-right panel of
Fig.~\ref{a1550} we present the radio contours of A1550 at 1.4 GHz
taken with the VLA in C configuration. This image has
been convolved to a resolution of
18\arcsec$\times$18\arcsec. 
The radio contours are overlaid on the optical
image taken from the SDSS. Three discrete radio
sources (labelled as A, B, and C) are clearly embedded in the radio halo. Spots of very faint
emission are also detected in the C configuration data set, but given
that they don't coincide with any discrete source detected in the
FIRST survey (Becker et al. 1995) we consider these emissions associated with the peaks of
the radio halo detected in the D configuration data set.

A total flux density of $\simeq$26.5$\pm$1.3 mJy is calculated 
from the D configuration image, by integrating the total
intensity surface brightness in the region of the diffuse emission
down to the 3$\sigma$ level. 
By subtracting the flux density  
of the discrete sources A, B, and C (see Table ~\ref{tab2}),
a flux density of 7.7$\pm$1.6 mJy appears to be associated 
with the low brightness diffuse emission. 
This flux density value corresponds to a radio power 
of $P_{\rm 1.4\,GHz} = 1.49\times10^{24}$\,{\rm W~Hz}$^{\rm -1}$.

In the top-right panel of Fig.~\ref{a1550} we present
the total intensity radio contours at 1.4 GHz with the 
VLA in D configuration after subtraction of discrete sources.
The flux density calculated on the image with the point sources 
subtracted is 8.2$\pm$0.5 mJy, while as measured from the 3$\sigma$ radio
isophote, the radio halo shows a LLS of about 6$'$ ($\simeq 1.41$ Mpc).

\subsection{CL 1446+26 - CL 1447+26 - ZwCl 1447.2+2619}

\begin{figure*}
\centering
\includegraphics[width=16 cm]{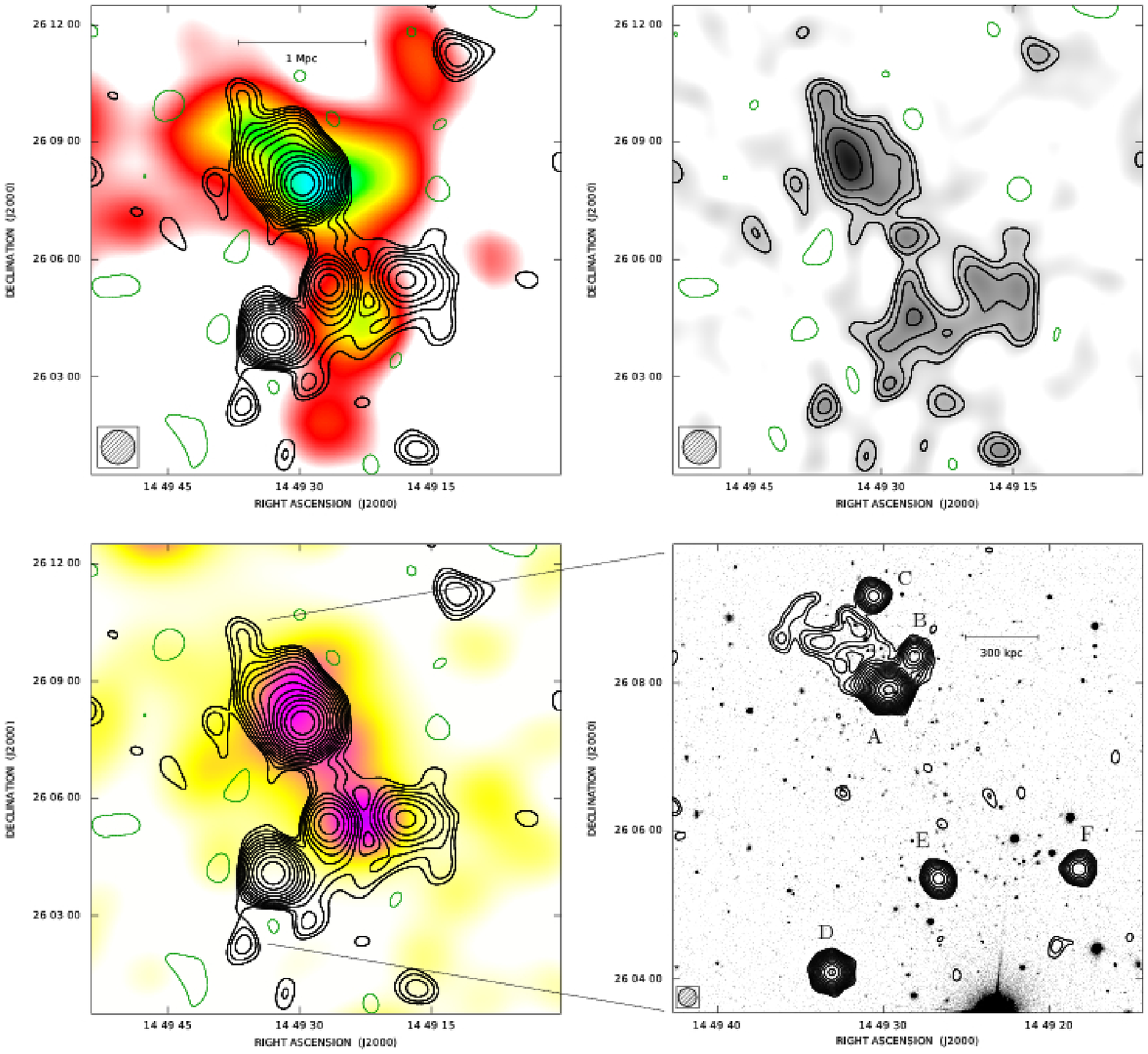}
\caption{{\bf CL 1446+26.} 
  {\it Top, left}: total intensity radio contours of CL 1446+26 at
  1.4 GHz with the VLA in D configuration.  The image has a FWHM of
  52\arcsec$\times$52\arcsec.  The contour levels are drawn 
  at $-$0.25, 0.25 mJy/beam, and the rest are spaced
  by a factor of $\sqrt2$.  The sensitivity (1-$\sigma$) is 0.08
  mJy/beam.  Total intensity radio contours are overlaid on the Rosat
  PSPC X-ray image in the 0.1$-$2.4 keV band, taken from the RASS.  
  The X-ray image has been convolved with a Gaussian of
  $\sigma$=45\arcsec.  
  {\it Top, right}: total intensity radio contours and grey scale image 
  at 1.4 GHz with the VLA in D configuration after subtraction of 
  discrete sources.
 {\it Bottom, left}: total intensity radio contours of
  CL 1446+26 overlaid on the isodensity map of likely cluster members.
 {\it Bottom, right}: zoom of total intensity radio contours in the center 
  of CL1446+26 at 1.4 GHz with the VLA in C configuration, image (VLA
  program AO149) taken from Giovannini et al. (2009).  The image has
  a FWHM of 15$''\times 15''$.  The first contour level is
  drawn at 0.15 mJy/beam, and the rest are spaced by a factor
  $\sqrt{2}$.  The sensitivity (1$\sigma$) is 0.05 mJy/beam.  The
  contours of the radio intensity are overlaid on the optical image
  taken from the SDSS (red plate).}
\label{cl1446}
\end{figure*}

CL 1446+26 is a rich galaxy cluster located at z=0.370 with an X-ray 
luminosity\footnote{The bolometric X-ray luminosity has been converted to the
0.1$-$2.4 keV band using Table 5
of B{\"o}hringer et al. (2004), assuming an
intra-cluster temperature of T$\simeq$ 5 keV.} 
in the 0.1$-$2.4 keV band 
of 3.41$\times$10$^{44}$erg/s (Wu et al. 1999).

At radio wavelengths, it has been studied by Owen et al. (1999) and Giovannini et al. (2009). 
The presence of a northern radio relic plus several radio galaxies 
were pointed out by Giovannini et al. (2009) 
(see also radio contours in the bottom-right panel of Fig.~\ref{cl1446}). 

Low resolution radio contours at 1.4 GHz of CL 1446+26 are shown in the top-left
and bottom-left panels of Fig.~\ref{cl1446}. This image was obtained with the VLA in D
configuration and has been convolved to a resolution of 52\arcsec$\times$52\arcsec.
At the top, the radio contours are overlaid on the cluster X-ray
image taken from the RASS. 
At the bottom, the radio contours are overlaid on the isodensity map of
likely cluster members.
The cluster is characterized by a disturbed morphology both in X-rays 
and in optical, with two main clumps, one to the North and one to the South.
Again, the optical and X-ray peaks are not coincident. 
As in A800 and A1550, the overall diffuse radio emission seems to follow 
more closely the galaxy than the intra-cluster density distribution.
Our lower resolution radio map shows in coincidence with the Northern   
clump a radio emission more extended than that 
obtained by Giovannini et al. (2009). In particular, this new image 
seems to indicate that
the diffuse emission is indeed a central radio halo rather than a peripheral
relic source, as previously stated. 
In addition, a new elongated diffuse radio emission 
in correspondence of the Southern sub-cluster has been detected. 
 
In the top-right panel of Fig.~\ref{cl1446} we present
the total intensity radio contours at 1.4 GHz with the 
VLA in D configuration after subtraction of discrete sources.
Given their location and morphology, in the following we define
the norther emission as a radio halo and the elongated southern emission
as a relic. The radio halo appears to be connected through a low brightness
diffuse bridge to the relic.
However, a deeper follow-up is necessary to clarify the nature of this system.

In the Northern emission a total flux density of 42.0$\pm$2.1 mJy 
is calculated from the D configuration image, by integrating the total intensity 
surface brightness in the region of the diffuse emission 
down to the 3$\sigma$ level. 
By subtracting the flux density 
of the discrete sources A, B and C (see Table\,\ref{tab2}) a flux density of
7.7$\pm$2.6 mJy appears to be associated to the radio halo. 
This flux density value corresponds to a radio
power of $P_{\rm 1.4\,GHz} = 3.57\times10^{24}$\,{\rm W~Hz}$^{\rm -1}$.
While in the Southern emission by subtracting the flux density
of the discrete sources D, E, and F from the total flux density 
(22.0$\pm$1.1 mJy), a flux density of
5.3$\pm$1.2 mJy appears to be associated to the relic. 
This flux density value corresponds to a radio
power of $P_{\rm 1.4\,GHz} = 2.46\times10^{24}$\,{\rm W~Hz}$^{\rm -1}$.

The LLS is about 4$'$ ($\simeq 1.22$ Mpc) and 5$'$ ($\simeq 1.53$ Mpc)
for the radio halo and relic, respectively. 
The flux densities calculated on the image with the point 
sources subtracted are consistent within the errors with the
values given in Table\,\ref{radio}, being 
 5.3$\pm$0.3 mJy in the halo and 5.5$\pm$0.3 mJy in the relic.

\section{Statistical properties}

\begin{table*}
\caption{Radio information of clusters containing diffuse emission.}
\label{radio}
\centering
\begin{tabular}{lclrrrrrr}
\hline\hline
Name    & z      & kpc/$''$ & Type  &$S_{1.4 GHz}$ & log $P_{1.4 GHz}$ & LLS & Radio$-$X-ray   & Radio$-$Optical \\
        &        &          &       &(mJy)       & (W/Hz)          & (Mpc) & Offset ($''$) & Offset ($''$)  \\
\hline
A800    & 0.2223 &  3.55    & H    &  10.6    &  $1.52\times10^{24}$ & 1.28  &  90 & 5 \\
A910    & 0.2055 &  3.34    & R   &  12.1    &  $1.45\times10^{24}$  & 1.00  & 135 & 200\\
A1550   & 0.2540 &  3.92    & H    &   7.7    &  $1.49\times10^{24}$ & 1.41  & 105 & 5   \\
CL1446+26N & 0.370 & 5.09 & H      &  7.7     &  $3.57\times10^{24}$ & 1.22  & 85  & 40  \\
CL1446+26S & 0.370 & 5.09 & R      &  5.3     &  $2.46\times10^{24}$ & 1.53  & 205 & 250 \\
\hline
\multicolumn{9}{l}{\scriptsize Col. 1: Cluster Name; Col. 2: Redshift; Col. 3: Angular to linear conversion; Col. 4: Type of diffuse radio emission:}\\
\multicolumn{9}{l}{\scriptsize H=halo, R=relic; Col. 5: Diffuse radio flux density at 1.4 GHz; Col. 6: Diffuse radio power at 1.4 GHz;}\\
\multicolumn{9}{l}{\scriptsize Col. 7: Radio Largest Linear Size (LLS); Col. 8: Radio-X-ray offset;}\\  
\multicolumn{9}{l}{\scriptsize Col. 9: Radio-Optical offset.}\\
\end{tabular}
\end{table*}

\begin{figure*}
\centering
\includegraphics[width=18 cm]{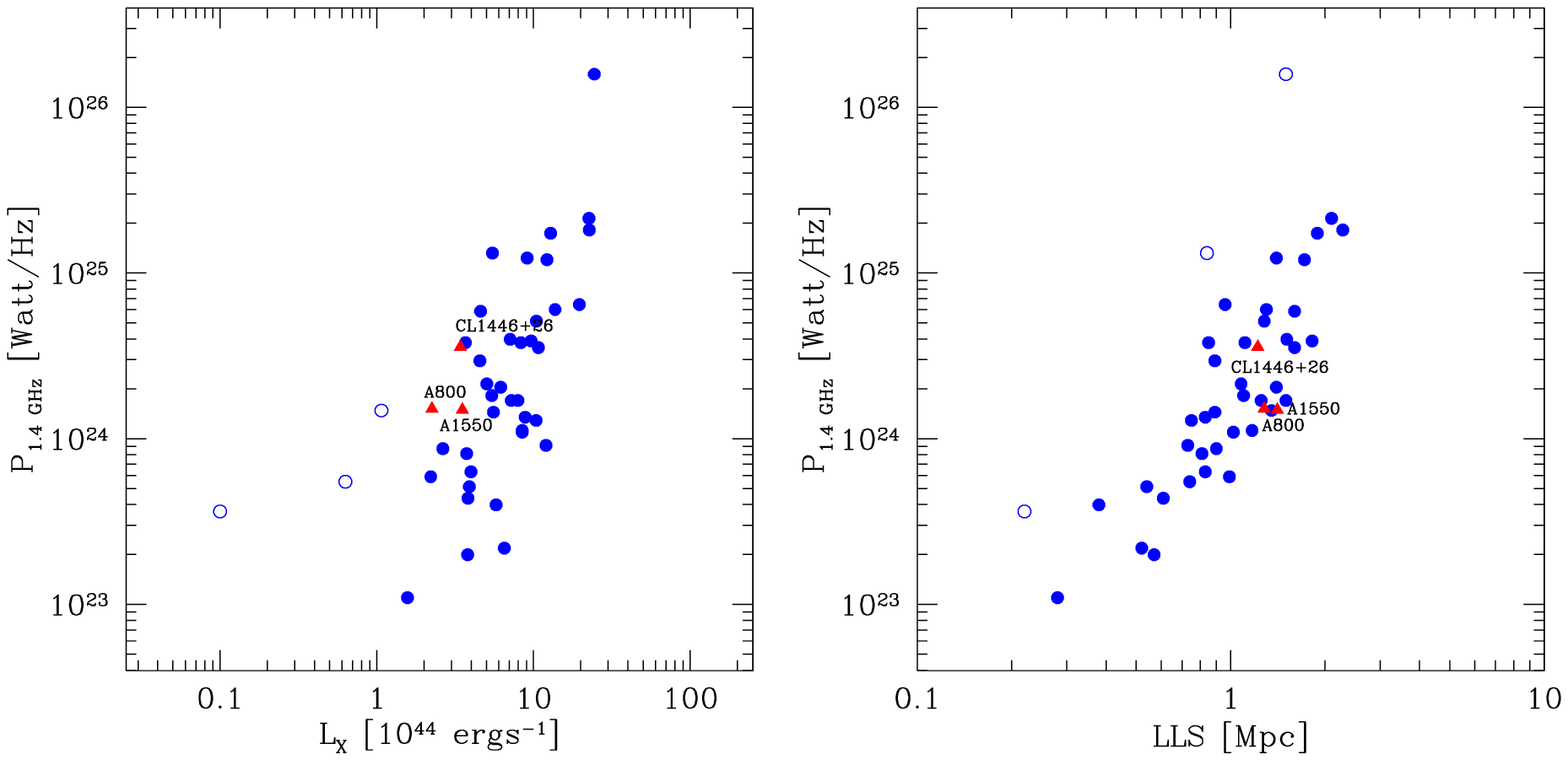}
\caption{
{\it Left}: Monochromatic radio power of radio halos at 1.4 GHz versus
the cluster X-ray luminosity in the 0.1$-$2.4 keV band. 
{\it Right}: Monochromatic radio power of halos at 1.4 GHz versus 
their largest linear size (LLS) measured at the same frequency.
The data are taken from the recent
compilation by Feretti et al. (2012) and 
references therein. Blue dots are classical radio halos while 
empty blue dots are outlier clusters. The new halos detected in this work 
are indicated with a red triangles.
}
\label{relation}
\end{figure*}

\begin{table}
\caption{Radio-X-ray offsets for a sample of halos.}
\centering
\begin{tabular}{lclrrl}
\hline\hline
Name      & z        & kpc/$''$ & Radio-X-ray    &  LLS &  Ref  \\
          &          &          & Offset ($''$) &  ($'$) &       \\
\hline
A209      &  0.2060  &   3.34   &   20     &  7.0 &  1 \\
A399      &   0.0718 &   1.35   &   135    &  7.0 &  2 \\ 
A401      &   0.0737 &   1.38   &    70    &  6.3 &  3 \\
A520      &   0.1990 &   3.25   &    20    &  5.7 &  4 \\
A523      &  0.1036  &  1.88    &      185 & 12.0 &  5 \\
A545      &  0.1540  &  2.64    &      50  &  5.6 &  3 \\
A665      &  0.1819  &   3.03   &      50  &   10 &  6 \\
A754      &  0.0542  &   1.04   &     142  & 15.8 & 3  \\ 
A773      &  0.2170  &   3.48   &      30  &  6.0 & 4  \\
A781      &  0.3004  &   4.42   &      55  &  6.0 & 7  \\
A800      &  0.2223  &   3.55   &      90  &  6.0 & *  \\
A851      &  0.4069  &   5.40   &      75  &  3.3 & 1  \\
A1213     &  0.0469  &   0.91   &     145  &  4.0 & 1  \\
A1351     &  0.3224  &   4.65   &     100  &  3.0 & 1  \\
A1550     &  0.2540  &   3.92   &     105  &  6.0 & *  \\ 
A1689     &  0.1832  &   3.05   &      15  &  4.0 & 8  \\
A1758a    &  0.2790  &   4.20   &      35  &  6.0 & 1  \\
A1914     &  0.1712  &   2.88   &      45  &  7.4 & 3  \\
A1995     &  0.3186  &   4.61   &      45  &  3.0 & 1  \\
A2034     &  0.1130  &   2.03   &      65  &  5.0 & 1  \\
A2163     &  0.2030  &   3.31   &      30  & 11.5 & 9  \\
A2218     &  0.1756  &   2.94   &      35  &  2.2  & 10  \\
A2219     &  0.2256  &   3.59   &      20  &  8.0 & 3  \\
A2254     &  0.1780  &   2.98   &      45  &  5.0 & 4  \\
A2255     &  0.0806  &   1.50   &      65  & 10.0 & 11  \\
A2294     &  0.1780  &   2.98   &      25  &  3.0 & 1  \\
A2319     &  0.0557  &   1.07   &      35  & 15.9 & 12 \\
A2744     &  0.3080  &   4.50   &      25  &  7.0 & 4  \\
RXCJ1314  &  0.2439  &   3.81   &     100  &  7.0 & 13 \\
CL1446+26 &  0.3700  &   5.09   &      85  &  4.0 & * \\
\hline
\multicolumn{6}{l}{\scriptsize Col. 1: Cluster Name; Col. 2: Redshift;}\\
\multicolumn{6}{l}{\scriptsize Col. 3: Angular to linear conversion; Col. 4: Radio - X-ray offset;}\\ 
\multicolumn{6}{l}{\scriptsize Col. 5: Radio largest linear size (LLS);}\\ 
\multicolumn{6}{l}{\scriptsize Col. 6: Reference to the radio halos at 1.4 GHz:}\\
\multicolumn{6}{l}{\scriptsize *=This work, 1=Giovannini et al. (2009),}\\ 
\multicolumn{6}{l}{\scriptsize 2=Murgia et al. (2010), 3=Bacchi et al. (2003),}\\
\multicolumn{6}{l}{\scriptsize 4=Govoni et al. (2001a), 5=Giovannini et al. (2011),}\\
\multicolumn{6}{l}{\scriptsize 6=Vacca et al. (2010), 7=Govoni et al. (2011),}\\ 
\multicolumn{6}{l}{\scriptsize 8=Vacca et al. (2011), 9=Feretti et al. (2001),}\\
\multicolumn{6}{l}{\scriptsize 10= Giovannini \& Feretti (2000), 11=Govoni et al. (2005),}\\
\multicolumn{6}{l}{\scriptsize 12=Feretti et al. (1997), 13= Feretti et al. (2005).}\\
\end{tabular}
\label{taboffset}
\end{table}


The basic properties of the new cluster 
diffuse radio sources (halos and relics) are reported in Table\,\ref{radio}.

The new relics presented here show properties similar to other published
relic sources (see e.g. the statistical characteristics of radio 
relics reviewed by Feretti et al. 2012), and are not further discussed 
in the following.

All the clusters presented in this paper and hosting new radio halos
(A800, A1550, CL1446+26) show a clear shift between the galaxies 
and gas distribution, as well as complex optical and X-ray morphologies. 
Joint numerical and observational analysis indicate that these 
are typical features of dynamically disturbed clusters, 
with a recent or ongoing major merging event (e.g. Maurogordato et al. 2011). 
Deeper X-ray observations and optical spectroscopic follow-up could help in 
re-constructing the collision scenario of these perturbed systems. 

In Fig.~\ref{relation} we plot the radio power calculated at 
1.4 GHz ($P_{\rm 1.4 GHz}$) versus the cluster X-ray
luminosity in the 0.1$-$2.4 keV band ($L_X$) and versus the
Largest Linear Size ($LLS$).
The blue dots refer to radio halos taken from the literature.
Among them the empty dots are outliers, i.e. 
on the left panel they refer to the few known 
radio halos A523 (Giovannini et al. 2011), A1213 (Giovannini et al. 2009), and 
0217+70 (Brown et al. 2011) that are over-luminous in radio with respect 
to the cluster X-ray luminosity, 
while on the right panel they refer to the few known
radio halos A1213, A1351 (Giovannini et al. 2009), 
MACS J0717.5+3745 (Bonafede et
al. 2009) which shows a size smaller than 
expected from their radio power.
The new radio halos are indicated with red triangles.  
They agree with both the $P_{\rm 1.4 GHz}-L_X$ and $P_{\rm 1.4 GHz}-LLS$ relations 
known for the other halos in clusters.

The observed correlation between the radio power of halos and their
size has been firstly found for giant radio halos
by Cassano et al. (2007). In a later analysis Murgia et al. (2009) by 
fitting the azimuthally averaged brightness profile with an exponential 
found that the radio halo emissivity is remarkably similar from one 
halo to the other, despite of quite different length-scales, 
in agreement with the $P_{\rm 1.4 GHz}-LLS$ correlation where larger halos show a 
higher radio power.

A close similarity between the radio and the X-ray structures
has been found in a number of clusters hosting a radio 
halo (see e.g. Govoni et al. 2001b).  
This similarity is generally valid for giant and regular halos. 
However, more irregular and asymmetric halos have
been found in the literature. In these halos, the radio emission 
may show significant displacement from the X-ray emission. Interestingly,
the new halos studied in this paper show a clear offset between
the radio and the X-ray peak (see Table\,\ref {radio}). In addition, 
the similarity between the radio and the X-ray morphology
is not clearly present and the radio 
emission seems to follow more closely the galaxy distribution rather than
the intra-cluster gas density distribution.

To investigate from a statistical point of view the offset between
the radio halo and the X-ray peak of the cluster emission,
we followed the analysis by Feretti et al. (2011).

\begin{figure}
\centering
\includegraphics[width=9 cm]{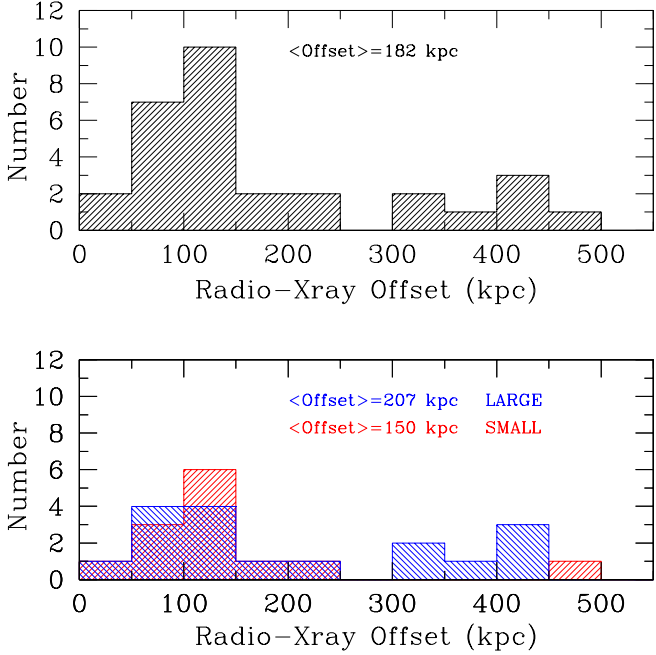}
\caption{ 
{\it Top}: Offset between the radio and X-ray peaks for a sample
of radio halos (see Table\,\ref{taboffset}). 
{\it Bottom}: Offset between the radio and X-ray peaks. 
The blue dashed area refers to giant halos ($LLS > 1$ Mpc) and
the red  dashed area refers to small halos ($LLS < 1$ Mpc), respectively.  
}
\label{offset1}
\end{figure}

In the top panel of Fig.~\ref{offset1} we present the distribution of
the offset between the peak position of radio halo
and the X-ray gas distribution for the new radio halos presented 
in this paper together with a sample of cluster at a redshift 
$z$$\leq$0.4 previously imaged at 1.4 GHz by our group 
(following a uniform data reduction strategy), for which good radio and 
X-ray data are available (see Table\,\ref{taboffset} for details).
The mean radio$-$X-ray offset is 182 kpc.
In the bottom panel of Fig.~\ref{offset1} the same distribution
refers to large halos with $LLS > 1$ Mpc (in blue) and small halos
with $LLS < 1$ Mpc (in red), respectively.
Both large and small radio halos can be significantly shifted, 
up to hundreds kpc, with respect to the X-ray center.  

\begin{figure}
\centering
\includegraphics[width=9 cm]{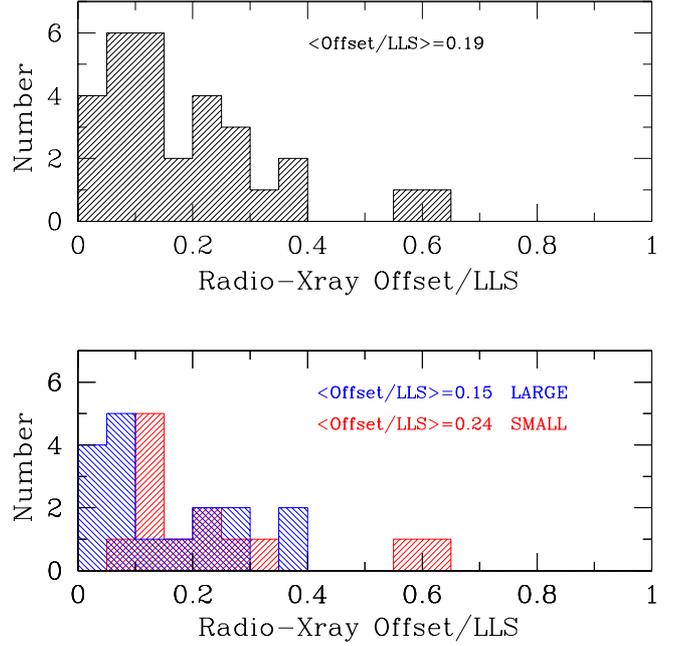}
\caption{ 
{\it Top}: Fractional offset (Offset/$LLS$) between the radio 
and X-ray peaks for a sample of radio halos (see Table\,\ref{taboffset}). 
{\it Bottom}: Fractional offset (Offset/$LLS$) between the 
radio and X-ray peaks. 
The blue dashed area refers to giant halos ($LLS > 1$ Mpc) and
the red  dashed area refers to small halos ($LLS < 1$ Mpc), respectively.  
}
\label{offset2}
\end{figure}

To highlight radio halos with the most pronounced asymmetric distribution,
the radio$-$X-ray offset normalized to the halo size ($LLS$) 
is plotted in Fig. \ref{offset2}. 
The mean value of the radio$-$X-ray offset normalized to the
$LLS$ is 0.19.
In the bottom panel of Fig.~\ref{offset2} the same distribution
is considered for large (in blue) and small (in red) halos separately. 
The mean value calculated for large halos is 0.15, while the
mean value calculated for small halos is 0.24. 
Thus we found that halos can be quite asymmetric with respect 
to the X-ray gas distribution, and this becomes more 
relevant when halos of smaller size are considered.

\section{Discussion}

Although the properties of the newly discovered, moderate-mass, 
radio halos are consistent with the $P_{\rm 1.4 GHz}-L_X$ and $P_{\rm 1.4 GHz}-LLS$ 
correlations known in the literature, 
their radio morphology does not matches exactly the thermal X-ray emission from
the intra-cluster medium.

By analyzing the radio X-ray offset statistically we found
a tendency for smaller halos to show larger displacements.
This result may have important consequences for the origin of radio halos, 
suggesting that the morphological properties of radio halos,
may be linked to the cluster dynamical state.

Major cluster mergers can supply energy to the radio emitting
particles, as well as amplify magnetic fields
(e.g. Roettiger et al. 1999, Dolag et al. 1999,  
Ricker \& Sarazin 2001, Dolag et al. 2005, Ryu et al. 2008, Vazza et al. 2009,
 Xu et al. 2010, Vazza et al. 2011, Xu et al. 2011).
Since magnetic fields have been found to be ubiquitous in galaxy clusters 
(e.g. Kronberg 1994, Carilli \& Taylor 2002, Clarke et al. 2001,
Govoni \& Feretti 2004, Bonafede et al. 2011), 
the crucial ingredient for the existence of diffuse radio halos
is the presence of synchrotron electrons with GeV energies.
Continuous acceleration of thermal electrons in galaxy clusters actually 
cannot work because the energy gained by the particles
is dissipated to the whole plasma on a timescale much shorter than that of the acceleration
process itself (see e.g. Petrosian 2001, Wolfe \& Melia 2006). 
On the other hand, it is believed that 
primary relativistic electrons are present in the cluster volume because they were
injected by AGN activity or by star-forming galaxies during the cluster dynamical history. 
This population of electrons suffers strong radiation losses mainly
because of synchrotron and inverse Compton emission, thus direct or
stochastic re-acceleration caused by merging shocks and turbulence
is needed to maintain their energy to the level necessary to produce the observed synchrotron
radio emission in the relatively weak intra-cluster magnetic fields 
(e.g. Jaffe 1977, Schlickeiser et al. 1987, Petrosian 2001,
Brunetti et al. 2001, Brunetti et al. 2011).
Alternatively, electrons can be continuously injected in the intra-cluster medium as secondary particles by
inelastic nuclear collisions between relativistic protons and nuclei in the thermal ambient intra-cluster medium (e.g. Dennison 1980, Kushnir et al. 2009,
Keshet \& Loeb 2010, En{\ss}lin et al. 2011).
The protons diffuse on large scale because their energy losses are negligible. 
They can continuously produce in situ electrons, distributed through the cluster volume. 
The possibility that the high
energy electrons, responsible for the synchrotron emission, arise from the decay of
secondary products of the neutralino annihilation in the dark matter halos of galaxy
clusters has also been suggested (Colafrancesco \& Mele 2001).

A detailed discussion of the origin of the non-thermal emission in 
galaxy clusters 
is beyond the aims of this work, which are in fact mainly observational. 
Nevertheless, we can explore, from a qualitative perspective, the reason
for the larger distortions and offsets observed among the smallest halos.
Cluster merger events are expected to release a significant amount of energy
in the intra-cluster medium. This energy is injected on large spatial scales, 
and then turbulent cascades may be generated. It is generally believed that
a fraction of this energy is finally delivered in relativistic 
particles acceleration and magnetic field amplification. 
Thus, in this framework, 
the observed offsets could be related to the age of the merger. 

In particular, a possible explanation for this behavior can be attributed to
the presence of intra-cluster magnetic field spatial variations on large scales
and/or a non-uniform distribution of relativistic electrons.
Indeed, on the basis of modeling, Vacca et al. (2010) found that if
the intra-cluster magnetic field and relativistic electrons fluctuations 
are close to the observing beam scale, the halo results smooth and rounded. 
Increasing the magnetic field and the relativistic electrons correlation length 
results in a much distorted radio halo morphology and in a significant 
offset of the radio halo peak from the cluster center.
Therefore, smaller halos may have a more distorted morphology
because they may be young systems in which the energy is still 
on large scales, leading to a non-uniform distribution of relativistic electrons
and/or a magnetic field correlation length 
larger than in the more extended, and dynamical older, radio halos.

\section{Conclusion}

In order to investigate the presence of extended diffuse radio sources 
in galaxy clusters, we studied the radio continuum emission
in a sample of galaxy clusters by analyzing archival VLA observations
at 1.4 GHz in D configuration.
These observations have been complemented by X-ray, optical, and higher 
resolution radio data. The higher resolution radio images ensure 
that the detected diffuse emission is real and
not due to a blend of discrete sources.
The point source subtraction is rather relevant for the 
detection of diffuse radio emission and the associated flux density.
Therefore, this procedure has been performed carefully by using
different methods.

These data permitted to increase the number of known large
scale diffuse radio emissions. 
Indeed, we discovered a new radio halo in the central region of A800 and A1550.
We detected a radio relic in the periphery of A910, and finally, 
we revealed both a halo and a relic in CL1446+26.

All the clusters presented in this paper and hosting new radio halos
show a clear shift between the galaxies 
and gas distribution as well as complex optical and X-ray morphologies,
typical features of dynamically disturbed clusters.
The radio X-ray similarity is not clearly present and the radio 
emission seems to follow more closely the galaxy distribution rather
than the intra-cluster gas
distribution. Interestingly, there is a clear offset between
the radio and the X-ray peak.

By analyzing this offset statistically 
we found that radio halos can be quite asymmetric with respect 
to the X-ray gas distribution, and this becomes more 
relevant when halos of smaller size are considered.
This behavior could be linked to the cluster dynamical state.
Indeed, Vacca et al. (2010) found that a much distorted radio halo 
morphology and a significant offset of the radio halo peak from 
the cluster center is expected increasing the magnetic field 
and the relativistic electrons correlation length in galaxy clusters.
Therefore, smaller halos may have a more distorted morphology
because they may be young systems in which the merger energy is still 
on large scales, leading to a non-uniform distribution of relativistic electrons
and/or a magnetic field correlation length 
larger than in the more extended, and dynamical older, radio halos.
Thus, in this framework, the observed offsets could be related 
to the age of the merger.

\begin{acknowledgements}
  We thank the referee for the suggestions that 
  improved the presentation of the paper.
  This research was partially supported by PRIN-INAF2009.  We
  acknowledge financial contribution from the agreement ASI-INAF
  I/009/10/0. CF and CB acknowledge financial support by the ``{\it Agence Nationale de la Recherche}'' 
  through grant ANR-09-JCJC-0001-01. The National Radio Astronomy Observatory (NRAO) is a
  facility of the National Science Foundation, operated under
  cooperative agreement by Associated Universities, Inc.  Funding for
  the SDSS and SDSS-II has been provided by the Alfred P. Sloan
  Foundation, the Participating Institutions, the National Science
  Foundation, the U.S. Department of Energy, the National Aeronautics
  and Space Administration, the Japanese Monbukagakusho, the Max
  Planck Society, and the Higher Education Funding Council for
  England. The SDSS Web Site is http://www.sdss.org/.  This research
  made use of Montage, funded by the National Aeronautics and Space
  Administration's Earth Science Technology Office, Computational
  Technnologies Project, under Cooperative Agreement Number NCC5-626
  between NASA and the California Institute of Technology. The code is
  maintained by the NASA/IPAC Infrared Science Archive.
\end{acknowledgements}

\appendix
\section{Images reliability}

\begin{figure*}[t]
\centering
\includegraphics[width=18 cm]{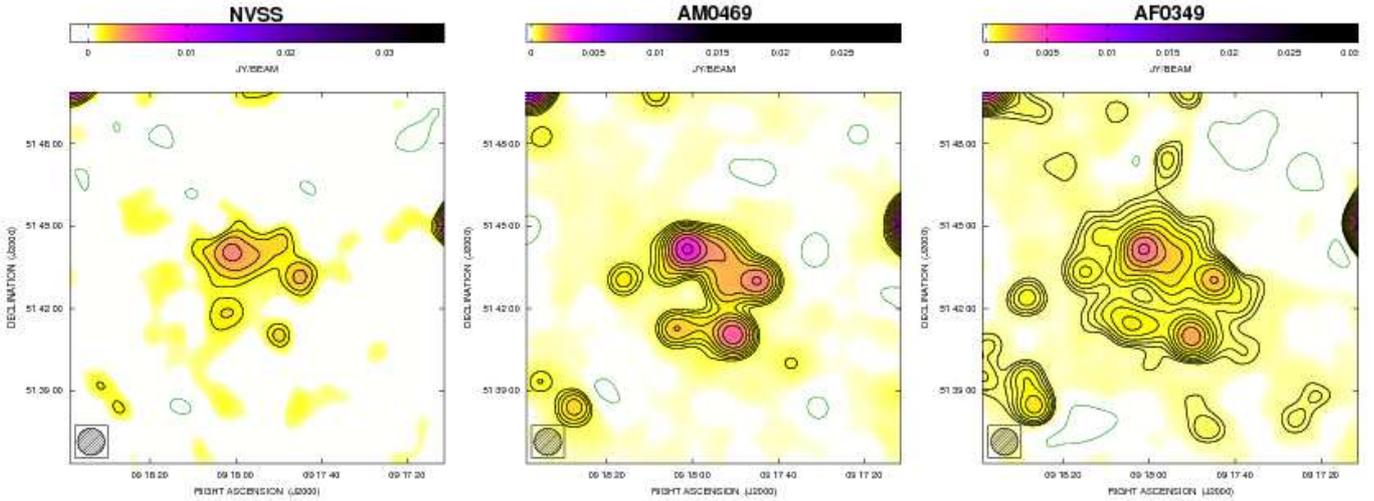}
\caption{
  Total intensity radio contours of A773 at 1.4 GHz obtained with 
  different VLA data sets. All images have
  been convolved to a FWHM resolution of 
  59\arcsec$\times$59\arcsec.
  {\it Left}: image form the NVSS (Condon et al. 1998).
  The contour levels are drawn at $-$1.2 (green), +1.2 mJy/beam, 
  and the rest are spaced by a factor of $\sqrt2$.
  {\it Middle}: image obtained from the AM0469 data set.
  The contour levels are drawn at $-$0.2 (green), +0.2 mJy/beam, 
  and the rest are spaced by a factor of $\sqrt2$.
  {\it Right}: image obtained from the AF0469 data set (Govoni et al. 2001a).
  The contour levels are drawn at $-$0.15 (green), +0.15 mJy/beam, 
  and the rest are spaced by a factor of $\sqrt2$.}
\label{a773}
\end{figure*}

In this work we made use of the archival VLA data set AM0469
to collect evidence of diffuse radio emission in galaxy clusters.
AM0469 contains about 40 clusters.
Some of them, like A773, were already known to host a radio halo.
To compare the quality of this data set with other
observations, in Fig. \ref{a773} we present the A773 field of view
as seen with the NRAO VLA Sky Survey (NVSS; Condon et al. 1998),
the AM0469 data set, and a deeper 
($\simeq$ 4 hours) VLA observation in C and D configurations 
(Govoni et al. 2001a, VLA program AF0349).
For a proper comparison images have been smoothed to a common resolution
of 59\arcsec$\times$59\arcsec.
We note that in the AM0469 data set we missed some diffuse cluster emission,
that is instead revealed in the deeper observation. However, this data set 
is fairly sensitive than the NVSS.
We remark that a significant breakthrough in the study of radio halos was 
obtained thanks to the NVSS (Giovannini et al. 1999).

\begin{thebibliography}{}

\bibitem{}
Abazajian, K. N., Adelman-McCarthy, J. K., Ag\"ueros, M. A., et al., 
2009,  ApJS, 182, 543

\bibitem{} 
Bacchi M., Feretti L., Giovannini G., Govoni, F., 2003, A\&A, 400, 465 

\bibitem{} 
Basu, K., 2012, \mnras, 421, L112 

\bibitem{}
Becker, R.~H., White, R.~L., \& Helfand, D.J., 1995, \apj, 450, 559

\bibitem{} 
B{\"o}hringer, H., Voges, W., Huchra, J.P., et al., 2000, \apjs, 129, 435 

\bibitem{} 
B{\"o}hringer, H., Schuecker, P., Guzzo, L., et al., 2004, A\&A, 425, 367 

\bibitem{} 
Bonafede, A., Feretti, L, Giovannini, G., et al., 2009, A\&A, 503, 707 

\bibitem{} 
Bonafede, A., Govoni, F., Feretti, L., et al., 2011, A\&A, 530, A24 

\bibitem{} 
Brown, S., \& Rudnick, L., 2011, \mnras, 412, 2 

\bibitem{} 
Brown, S., Duesterhoeft, J., Rudnick, L., 2011, \apjl, 727, L25 

\bibitem{} 
Brunetti, G., Setti, G., Feretti, L., \& Giovannini, G.\ 2001, \mnras, 320, 365 

\bibitem{} 
Brunetti, G.\ 2011, \memsai, 82, 515

\bibitem{} 
Buote, D.A., 2001, \apjl, 553, L15 

\bibitem{}
Carilli, C.L., \&  Taylor, G.B., 2002, ARA\&A 40, 319 

\bibitem{} 
Cassano, R., Brunetti, G., Setti, G., Govoni, F., Dolag, K., 2007, \mnras, 378, 1565 

\bibitem{} 
Cassano, R., Ettori, S., Giacintucci, S., et al., 2010, \apjl, 721, L82 

\bibitem{} 
Clarke, T.~E., Kronberg, P.~P., B{\"o}hringer, H., 2001, \apjl, 547, L111

\bibitem{} 
Colafrancesco, S., 1999, Diffuse Thermal and Relativistic Plasma in Galaxy Clusters, 269 

\bibitem{} 
Colafrancesco, S., \& Mele, B.\ 2001, \apj, 562, 24

\bibitem{}
Condon J.J., Cotton W.D., Greisen E.W., et al., 1998, AJ 115, 1693

\bibitem{} 
Dennison, B.\ 1980, \apjl, 239, L93

\bibitem{} 
Dolag, K., Bartelmann, M., Lesch, H., 1999, \aap, 348, 351

\bibitem{} 
Dolag, K., Vazza, F., 
Brunetti, G., \& Tormen, G., 2005, \mnras, 364, 753 

\bibitem{} 
Ebeling, H., Edge, A.C., Allen, S.W., et al., 2000, \mnras, 318, 333 

\bibitem{} 
En{\ss}lin, T., Pfrommer, C., Miniati, F., \& Subramanian, K.\ 2011, \aap, 527, A99

\bibitem{} 
Feretti, L., Giovannini, G., Bohringer, H., 1997, New Astronomy , 2, 501 

\bibitem{} 
Feretti, L., Fusco-Femiano, R., Giovannini, G., Govoni, F., 2001, \aap, 373, 106

\bibitem{} 
Feretti, L., 2002, The Universe at Low Radio Frequencies, 199, 133 

\bibitem{} 
Feretti, L., Schuecker, P., B{\"o}hringer, H., Govoni, F., Giovannini, G.,
2005, \aap, 444, 157 

\bibitem{} 
Feretti, L., Giovannini, G., Govoni, F., Murgia, M., 2011, 
IAU Symposium, 274, 340 

\bibitem{} 
Feretti, L., Giovannini, G., Govoni, F., Murgia, M., 2012,
Astronomy and Astrophysics Review in press, arXiv:1205.1919

\bibitem{} 
Ferrari, C., Benoist, C., Maurogordato, S., Cappi, A., Slezak, E., 2005, A\&A, 430, 19

\bibitem{} 
Ferrari, C., Govoni, F., Schindler, S., Bykov, 
A.M., Rephaeli, Y., 2008, Space Science Reviews, 134, 93 

\bibitem{} 
Giacintucci, S., Venturi, T., Cassano, R., et al., 2009 \apjl, 704, L54 

\bibitem{} 
Giovannini, G., Tordi, M., Feretti, L., 1999, \na, 4, 141 

\bibitem{} 
Giovannini, G., \& Feretti, L., 2000, New Astronomy, 5, 335 

\bibitem{} 
Giovannini, G., Bonafede, A., Feretti, L., et al., 2009, A\&A, 507, 1257  

\bibitem{} 
Giovannini, G., Feretti, L., Girardi, M., et al., 2011, \aap, 530, L5 

\bibitem{} 
Goto, T., Sekiguchi, M., Nichol, R.C., et al., 2002, AJ, 123, 1807 

\bibitem{} 
Govoni, F., Feretti, L., Giovannini, G., et al., 2001a, \aap, 376, 803 

\bibitem{} 
Govoni, F., En{\ss}lin, T.~A., Feretti, L., Giovannini, G., 2001b, \aap, 369, 441 

\bibitem{}
Govoni, F., Feretti, L., 2004, Int. J. Mod. Phys. D, Vol. 13, N.8, p. 1549

\bibitem{} 
Govoni, F., Markevitch, M., Vikhlinin, A., et al.,\ 2004, \apj, 605, 695 

\bibitem{} 
Govoni, F., Murgia, M., Feretti, L., et al.,\ 2005, \aap, 430, L5 

\bibitem{} 
Govoni, F., Murgia, M., Feretti, L., et al.\ 2006, \aap, 460, 425 

\bibitem{} 
Govoni, F., Murgia, M., Giovannini, G., Vacca, V., Bonafede, A.,
2011, \aap, 529, A69 

\bibitem{} 
Greisen, E.W., Spekkens, K., van Moorsel, G.A., 2009, \aj, 137, 4718

\bibitem{}
Hambly, N.C., MacGillivray, H.T., Read, M.A., et al., 2001, MNRAS, 326, 1279

\bibitem{} 
Jaffe, W.~J., 1977, \apj, 212, 1

\bibitem{} 
Kempner, J.C., \& Sarazin, C.L., 2001, \apj, 548, 639 

\bibitem{} 
Keshet, U., \& Loeb, A., 2010, \apj, 722, 737 

\bibitem{}
Kronberg, P.P., 1994, Rep. Progr. Phys., Vol 57, 325 

\bibitem{} 
Kushnir, D., Katz, B., \& Waxman, E.\ 2009, \jcap, 9, 24

\bibitem{} 
Liang, H., 1999, Diffuse Thermal and Relativistic Plasma in Galaxy Clusters, 33 
 
\bibitem{}
Maurogordato, S., Sauvageot, J.L., Bourdin, H., et al., 2011, A\&A, 525, 79
 
\bibitem{} 
Murgia, M., Govoni, F., Feretti, L., et al., 2004, \aap, 424, 429 

\bibitem{} 
Murgia, M., Govoni, F., Markevitch, M., et al., 2009, \aap, 499, 679 

\bibitem{} 
Murgia, M., Govoni, F., Feretti, L., Giovannini, G., 2010, \aap, 509, A86 

\bibitem{} 
Orr{\'u}, E., Murgia, M., Feretti, L., et al., 2007, \aap, 467, 943 

\bibitem{} 
Owen, F., Morrison, G., Voges, W., 1999, 
Diffuse Thermal and Relativistic Plasma in Galaxy Clusters, 9 

\bibitem{} 
Petrosian, V., 2001, \apj, 557, 560

\bibitem{} 
Pizzo, R.F., de Bruyn, A.G., Bernardi, G., Brentjens, M.A., 2011, A\&A, 525, A104 

\bibitem{} 
Ricker, P.M., \& Sarazin, C.L., 2001, \apj, 561, 621 

\bibitem{} 
Roettiger, K., Stone, J.M., Burns, J.O., 1999, \apj, 518, 594 

\bibitem{} 
Rossetti, M., Eckert, D., Cavalleri, B.M., et al., 2011, \aap, 532, A123 

\bibitem{} 
Rudnick, L., \& Lemmerman, J.A., 2009, \apj, 697, 1341 

\bibitem{} 
Ryu, D., Kang, H., Cho, J., \& Das, S., 2008, Science, 320, 909

\bibitem{} 
Schlickeiser, R., Sievers, A., \& Thiemann, H.\ 1987, \aap, 182, 21

\bibitem{} 
Schuecker, P., B{\"o}hringer, H., Reiprich, T.H.,  
Feretti, L., 2001, \aap, 378, 408 

\bibitem{} 
Tribble, P.C., 1991, \mnras, 253, 147 

\bibitem{} 
van Weeren, R.J., R{\"o}ttgering, H.J.A., Br{\"u}ggen, M., Cohen, A.,
2009, \aap, 508, 75 

\bibitem{} 
van Weeren, R.J., Br{\"u}ggen, M., R{\"o}ttgering, H.J.A., et al.,
2011, \aap, 533, A35 

\bibitem{} 
Vacca, V., Murgia, M., Govoni, F., et al., 2010, \aap, 514, A71

\bibitem{} 
Vazza, F., Brunetti, G., Kritsuk, A., et al., 2009, \aap, 504, 33 

\bibitem{} 
Vazza, F., Brunetti, G., Gheller, C., Brunino, R., Br{\"u}ggen, M.,
2011, \aap, 529, A17 

\bibitem{} 
Venturi, T., Giacintucci, S., Brunetti, G., et al., 2007, \aap, 463, 937 

\bibitem{} 
Venturi, T., Giacintucci, S., Dallacasa, D., et al., 2008, \aap, 484, 327 

\bibitem{} 
Xu, H., Li, H., Collins, 
D.C., Li, S., Norman, M.L., 2010, \apj, 725, 2152 

\bibitem{} 
Xu, H., Li, H., Collins, 
D.C., Li, S., Norman, M.L., 2011, \apj, 739, 77 

\bibitem{} 
Wolfe, B., \& Melia, F.\ 2006, \apj, 638, 125 

\bibitem{} 
Wu, X.-P., Xue, Y.-J., Fang, L.-Z., 1999, \apj, 524, 22 

\end{thebibliography}
\end{document}